# Autonomous Mars Rover Module for Soil Sampling and Life Component Analysis

Bibek Adhikari*, Rishab Rijal†, Rakesh Yadav ‡, Nikchey Khatri§, and Sandesh Dhakal¶
*The University of Texas at Arlington, Arlington, TX 76019*

**The search for extraterrestrial life has long been a primary focus of scientific exploration, driven by rapid advancements in technology and our understanding of the universe. The discovery of water on Mars has sparked significant interest, raising the question of whether life could exist on the planet. This study proposes a novel approach to simulate and illustrate the detection of life using a proof-of-life module integrated into a Mars rover. The module is an autonomous system capable of traveling to designated regions, excavating soil, collecting samples, and performing biochemical testing onboard the rover itself. The project is inherently multidisciplinary, integrating mechanical systems such as a drill mechanism, and a vacuum system, alongside biochemical analysis for soil testing. The module is capable of successfully detecting the presence or absence of living components of life from the collected soil particles. This proof-of-life module serves as a proof-of-concept for autonomous life detection in extraterrestrial environments and lays the foundation for future exploration missions.**

## I. Introduction

Human curiosity has been the driving force behind countless discoveries throughout history, from the discovery of fire and electricity to the understanding of microscopic structures like atoms and DNA. This innate curiosity has also pushed humanity to explore beyond our planet, leading to inquiries about the mechanisms governing the universe and the possibility of life on other worlds. Among these endeavors, Mars, the "Red Planet," has long fascinated scientists and the general public alike. Its striking red hue, visible even to the naked eye, has sparked wonder for centuries, and the invention of the telescope in the late 1800s allowed humans to observe Mars with greater clarity. Extensive research has revealed that Mars shares several similarities with Earth [1–3]. Geological evidence suggests that the planet once had flowing water, potentially even vast oceans, before a catastrophic event transformed its surface into a barren landscape dominated by iron oxide, carbon dioxide, and dust [4]. The discovery of ancient water bodies strongly indicates that Mars may have once been capable of supporting life. This tantalizing prospect has motivated scientific missions to study the Martian surface and search for traces of past or present life.

To further this exploration, rover, semi-autonomous vehicles designed to traverse and analyze extraterrestrial terrains, have been deployed to Mars. To date, six rovers have been sent to the planet, three of which remain operational. The rovers are crucial in extraterrestrial lands for the multi-purposes such as crater and resource identification [5], mapping of the environment [6], and several other researches [7]. While these missions have yielded numerous discoveries, direct evidence of life on Mars has yet to be found [8]. This ongoing mystery underscores the need for innovative approaches to planetary exploration and life detection. The paper presents a method to develop and test a novel life-detection module as part of a rover named Discovery. The proof of concept module integrates soil sampling and onboard biochemical analysis to detect residues indicative of life. The module employs cutting-edge mechanisms for collecting and testing soil particles, aiming to enhance the capabilities of current rover technologies. By bridging the gap between soil sampling and real-time biochemical analysis, this module could serve as a foundational tool for future Mars exploration missions.

The rest of the paper is organized as follows: Section II formulates the problem of the paper. Section III describes the methodology for the design and analysis of the Mars rover module. Section IV presents the results and findings of the study. Finally, Section V concludes with a discussion of the outcomes and insights gained from the soil sampling and life-detection experiments performed by the Mars rover module.

---

*Graduate student, Department of Mechanical and Aerospace Engineering
†Ph.D. student, Department of Mechanical and Aerospace Engineering
‡Ph.D. student, Department of Mechanical and Aerospace Engineering
§Graduate student, Department of Mechanical and Aerospace Engineering
¶Graduate student, Department of Mechanical and Aerospace Engineering

## II. Problem Formulation

This project aims to design and develop a life detection module for the MARS rover, capable of collecting soil samples from designated locations on the Martian surface and analyzing them for signs of life, both extant and extinct. The life detection module is designed to be mounted on the rover's chassis, where it can perform real-time analysis of soil samples. Once collected, the module will conduct onboard testing to detect any biological components within the sample, such as microbial life or organic molecules that could suggest the presence of life. Upon identifying potential signs of life, the module will transmit the results to the rover's ground station for further analysis and interpretation. To ensure reliability and functionality, the module is first tested in an Earth-based environment that replicates Martian conditions. The module is built to compete in the University Rover Challenge (URC) in the desert of Utah. The URC is the world's foremost robotics competition for designing the next generation rover for Mars exploration. In the competition, the rover should be able to participate on three distinct missions, and the proof of life is one of them.

## III. Methodology

### A. Rover Module

The proof of life module, integrated into the Mars Rover Discovery, was developed to collect soil samples from the Martian surface and analyze them for biochemical markers indicative of life. The module's aluminum alloy frame was chosen for its lightweight, robust, and machinable properties. Inside, multiple subsystems work together seamlessly to achieve the mission's objectives, including a pulley mechanism for vertical translation of the module, a drill subsystem for soil collection, a vacuum subsystem for sample transportation, and a biochemical testing unit for life detection analysis. The top plate of the module is used for vacuum housing and vacuum holder section is present in the rear side of the module plate as seen in Figure 1. The bottom plate of the module system consists of suction cup housing section. The hole in the bottom plate helps in suction cup and drill translation to the ground. The design emphasized ease of assembly, modularity, and reliable performance in a simulated Martian environment. The module's position atop the rover ensured efficient functionality and protected sensitive components during operation.

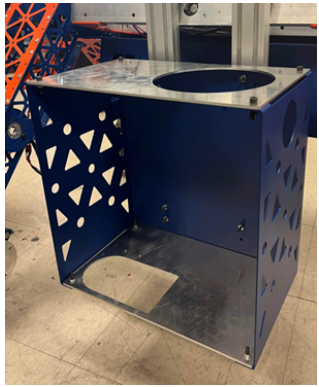

**Figure 1. Proof of life module housing**

### B. Pulley Subsystem

The pulley system is a key component of the proof of life module, responsible for its vertical translation. Powered by a high-torque 12V DC motor, the system operates by rotating a nylon rope via a spool, enabling the module to move upward or downward. The pulley mechanism was chosen over a lead screw system for its simplicity, cost-effectiveness, and ease of manufacturing. The design includes a motor shaft housing that supports the spool, allowing smooth rotation. The spool, 3D printed from ABS plastic, features a passage for the nylon rope, which adjusts its length during operation. An encoder mount tracks the module's position and ensures accurate movement, while a motor mount stabilizes the motor, minimizing vibrations for efficient operation. A double roller mount ensures the smooth passage of the nylon rope, maintaining tension and reducing friction. The entire system is integrated into the rover's chassis, with a rail assembly providing additional stability. The pulley components, including the spool and mounts, are lightweight and durable, ensuring reliable performance. This mechanism efficiently enables vertical translation, making it a cost-effective and robust solution for the module's functionality. The CAD and the physical hardware of the system is shown in Figure 2.

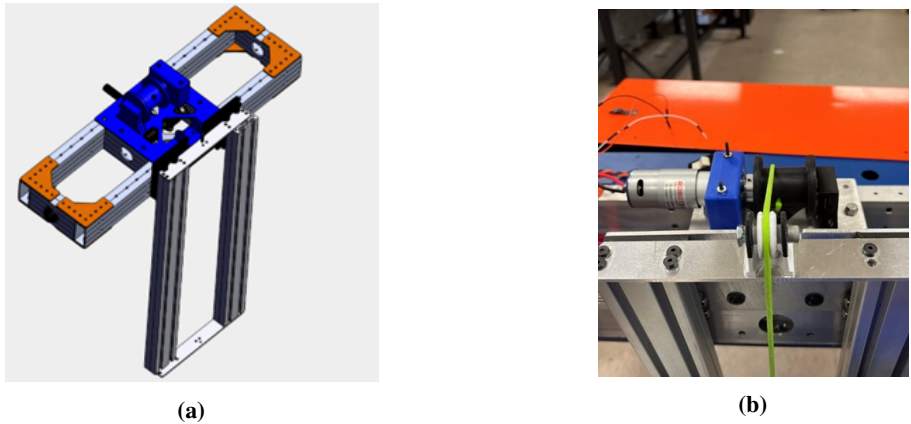

(a)          (b)

**Figure 2.** Pulley system CAD model (on left column) and physical hardware (on right column)

### C. Drill Subsystem

The drill system is designed to dig and collect soil samples for analysis. It incorporates a 785-gear rack kit assembly with two vertical beams featuring teeth on their inner surfaces. The gear rack assembly also includes a steel motor case housing a Hitec HS-788HB brushless DC motor. As the servo gear moves along the vertical teeth of the beams, the motor case translates vertically. The beams are securely attached to the module's top and bottom plates, ensuring stability during operation. A rectangular plate made of 6061 aluminum is attached to the U-channel of the motor case, providing structural support for the drill assembly. A 12V DC motor drives a 3.5-inch spiral flute drill bit made of structural steel, capable of digging to a maximum depth of 1.88 inches. The drill bit, with a 0.25-inch shank, is coupled to the motor, while a 3D-printed PLA motor case secures the motor and drill in place. This motor case is affixed to the aluminum plate with a screw and is further integrated with a suction cup concentric to the drill bit, enabling soil collection through the vacuum system. The suction cup, also 3D-printed from ABS plastic, houses the drill bit and the vacuum pipe, facilitating efficient soil suction. As the drill rotates and translates simultaneously, soil is loosened and transported to the vacuum system for collection. The system is powered by a 12V DC motor for rotation, while vertical translation is achieved via the gear rack kit's servo motor. The CAD and the physical hardware of the system is shown in Figure 3.

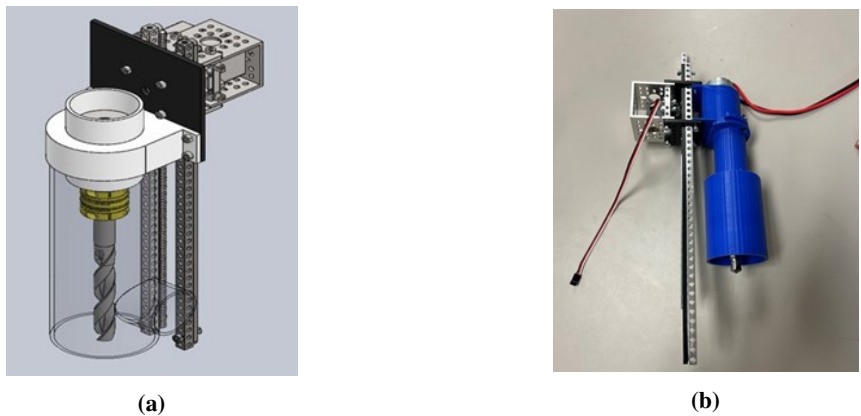

(a)          (b)

**Figure 3.** Drill system CAD model (on left column) and physical hardware (on right column)

Analysis of the drill system using ANSYS Workbench revealed that the motor case experiences minimal deformation during operation, with a maximum deformation magnitude of 47.563 µin at the motor shaft base. The aluminum plate provides a stable anchor point with negligible deformation. The system's factor of safety, calculated to be 15, confirms its structural integrity under operational loads. Overall, the drill system demonstrated reliability and precision in soil collection for further biochemical analysis.

### D. Vacuum Subsystem

The vacuum system is based on the principle of centrifugal pumping, where a low-pressure zone is created at the center of a rotating impeller. This pressure difference draws air from the surrounding high-pressure environment, carrying soil particles along with it into the vacuum container. Inside the container, an air filter separates the soil sample from the airflow. The filtered air exits through a diffuser outlet, while the collected soil is deposited into a testing bin located at the bottom of the container. The CAD and the physical hardware of the vacuum system is shown in Figure 4, which consist of vacuum housing at the bottom covered by the top plate at the top. Between the housing and the plate, there are impeller and diffuser. The impeller, shown in Figure 5, the rotating component of the system, is responsible for generating the centrifugal force that drives the airflow. As the impeller rotates, it accelerates the air radially outward, creating the necessary low-pressure zone at its center. This impeller, manufactured using ABS plastic and 3D printing, is powered by a 12V DC motor capable of operating at 10,000 RPM. The motor's specifications, favoring high rotational speed and low torque, ensure efficient air acceleration and soil suction. The diffuser complements the impeller by managing the airflow and increasing air pressure as it moves radially outward. The diffuser's design reduces the velocity of the air while enhancing its pressure, ensuring consistent and controlled unidirectional airflow from the inlet to the outlet. The diffuser also incorporates an air filter at its inlet, which separates soil particles from the incoming air. Once the vacuum operation stops, the collected soil settles into the testing bin through the bottom channel of the vacuum container.

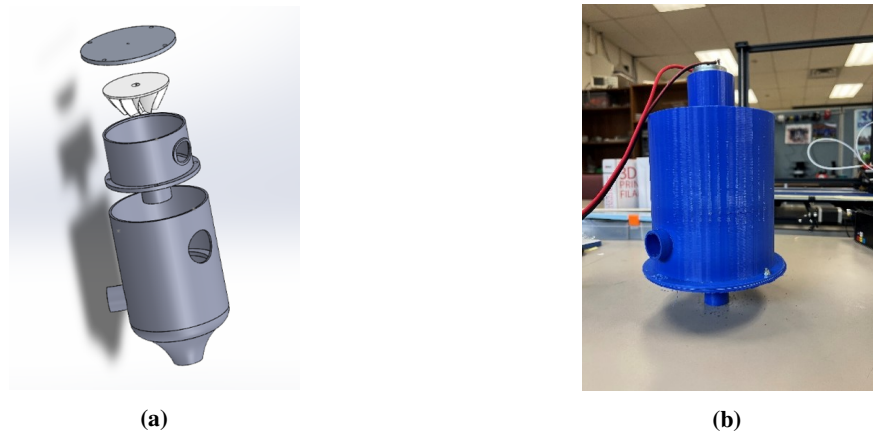

(a)      (b)

**Figure 4. Vacuum system CAD model (on left column) and physical hardware (on right column)**

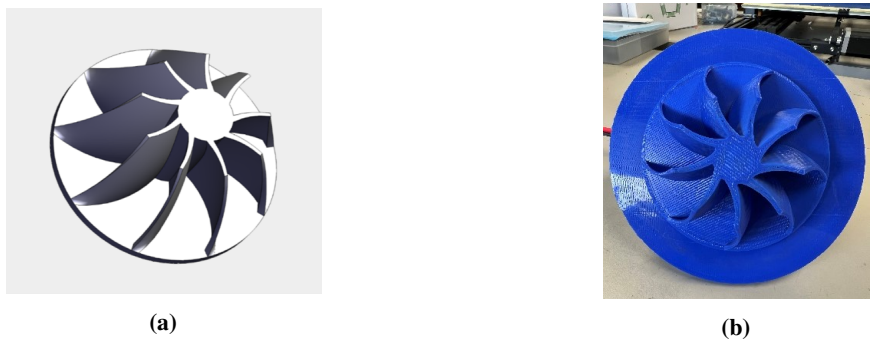

(a)      (b)

**Figure 5. Vacuum impeller CAD model (on left column) and physical hardware (on right column)**

To validate the design, Computational Fluid Dynamics (CFD) analysis was conducted using ANSYS Fluent. The velocity profiles demonstrated that the airflow accelerates toward the diffuser outlet, confirming the impeller's ability to efficiently transport soil-laden air, as seen in Figure 6a. Similarly, pressure profiles indicated a clear pressure gradient, with the lowest pressure at the center of the impeller and increasing radially outward, as seen in Figure 6b. This validated the vacuum system's capability to create the required suction and maintain effective soil transport. Following validation, the vacuum system components were manufactured entirely using 3D printing with ABS plastic. The system was powered by a 12V supply from the rover and controlled via Arduino, ensuring seamless integration with other subsystems. Laboratory tests confirmed the vacuum system's functionality, demonstrating reliable soil collection and deposition into the testing bin.

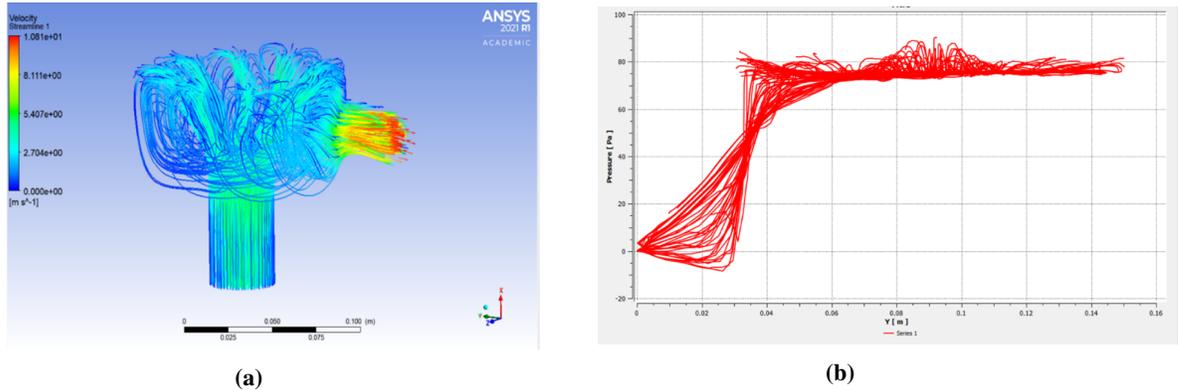

(a)  (b)

**Figure 6. Velocity profile in the diffuser (on the left column) and pressure distribution along the radial direction of the impeller (on the right column)**

### E. Sample Collection and Testing Subsystem

The testing bin, situated at the module's base, consists of three separate chambers equipped with transparent beakers for biochemical analysis. Each chamber can be rotated into position using a 12V DC motor, enabling sequential tests at different locations. The beakers are pre-filled with reagents for protein and carbohydrate testing. The bin's rotation mechanism, supported by stainless steel ball bearings, minimizes friction and ensures smooth operation. All components of the bin, except for the glass beakers, were 3D printed using ABS plastic for rigidity and ease of manufacturing.

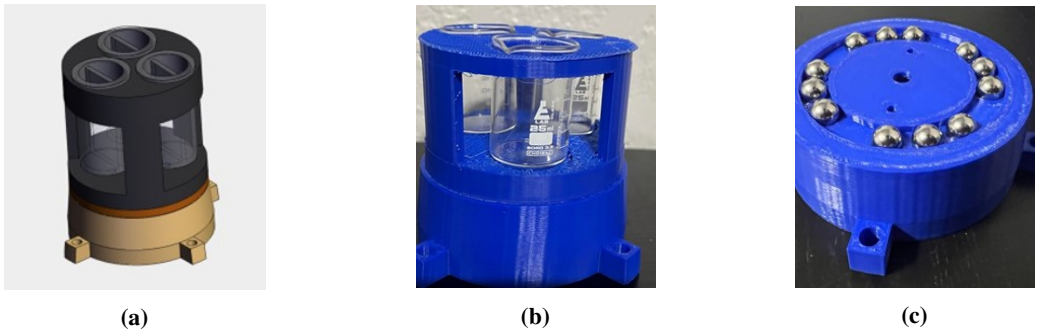

(a)  (b)  (c)

**Figure 7. Sample collection bin CAD model (on left column), physical hardware (on middle column) and bin base with bearing (on right column)**

The collected soil samples were tested for the presence of life by analyzing two primary biochemical components: proteins and carbohydrates. These components are essential indicators of life. The testing process was based on color changes in reagent solutions upon reaction with the soil samples, which were monitored using a camera mounted inside the module. Tests for lipids and nucleic acids were excluded due to technical and design limitations. The lipid test, using Sudan IV, produced indistinct color changes that were difficult to capture with the camera in a heterogeneous solution. Similarly, nucleic acid testing was deemed too complex and expensive for the scope of this module.

**F. Electronics and Programming**

The circuit layout for the proof of life module integrates all components to ensure seamless operation. Power and ground connections are denoted by red and black lines in the layout, while signal wires are represented by various other colors. Figure 8 outlines the circuit layout of the overall system, which consists an Arduino Mega 2560 micro-controller, which serves as the central control unit and is powered by a 5V supply from the rover's power outlet. The ultrasonic sensor is directly connected to the Arduino, with its voltage pin connected to the 5V pin on the board, the echo pin connected to digital pin 2, the trig pin to digital pin 3, and its ground pin to the Arduino's ground. The brushless motor, which operates using a servo signal, is also connected to the Arduino. Its signal wire is linked to the PWM input pin (pin 4), and it is supplied with 5V power.

For controlling the 12V DC motors that drive the pulley, drill, vacuum, and testing bin mechanisms, two H-bridge motor drivers (H-bridge A and H-bridge B) are utilized. Both H-bridges receive 12V power from the rover's power outlet, with a common ground established between the H-bridges and the Arduino. The pulley motor is connected to channel A of H-bridge A, while the drill motor is connected to channel B of the same H-bridge. Similarly, the vacuum motor is connected to channel A of H-bridge B, and the bin motor is connected to channel B. The polarity of the motors is controlled by the input pins of the H-bridges, with IN1 and IN2 controlling channel A, and IN3 and IN4 controlling channel B for each H-bridge. The ENA and ENB pins on the H-bridges are used to regulate motor speeds for the respective channels. The Arduino is programmed to operate the motors sequentially. First, the pulley motor connected to channel A of H-bridge A is activated, followed by the drill motor on channel B of H-bridge A. Afterward, the vacuum motor on channel A of H-bridge B is engaged, and finally, the bin motor on channel B of the same H-bridge is activated. This sequence ensures that the module's operations, from soil collection to testing, proceed smoothly and automatically.

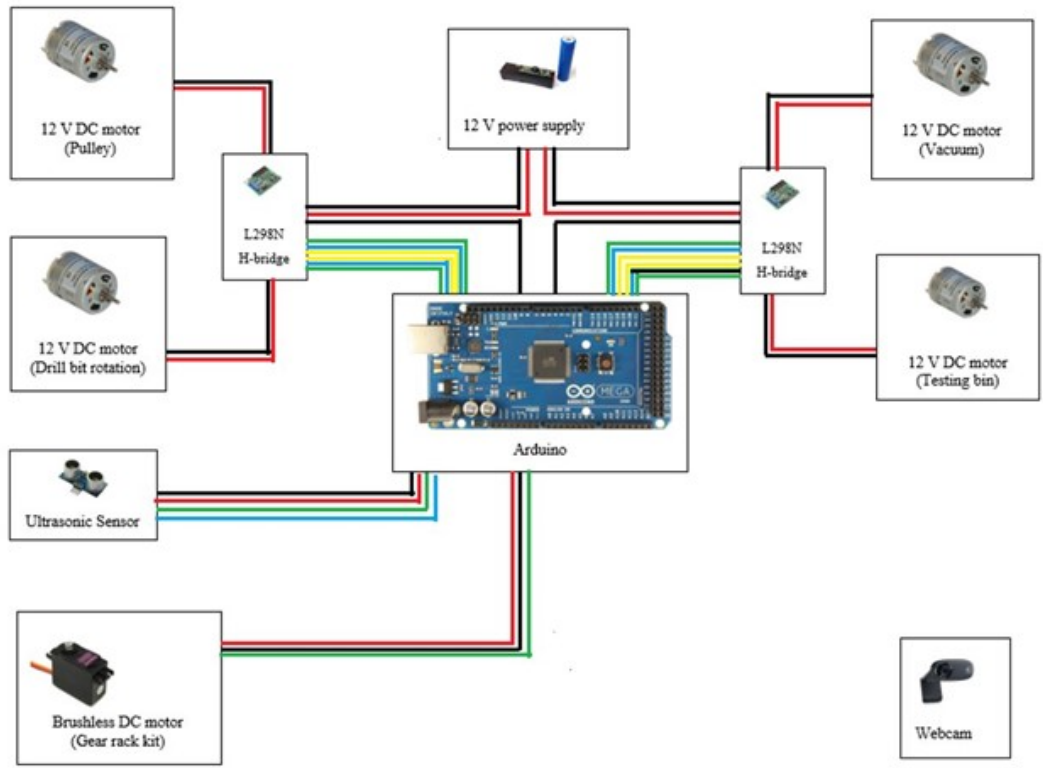

**Figure 8. Circuit layout of the module**

**G. Integrated Mechanism**

The system operates autonomously based on a programmed sequence controlled by an Arduino. The process begins with the vertical translation of the module, driven by the pulley mechanism. Once the module is stabilized, an ultrasonic sensor measures the distance between the module's base and the ground. This distance is used to guide the gear rack kit system, enabling the drill to lower and make contact with the ground. Once in position, the drill starts rotating to excavate the soil. The loosened soil is then collected by the vacuum system, which creates a low-pressure zone using a rotating impeller. The airflow generated by the impeller transports the soil into the vacuum container. Inside the container, an air filter separates the soil particles from the air, which exits through the diffuser outlet. The soil, now filtered, is deposited at the bottom of the vacuum container. After the vacuum mechanism completes its operation, the soil sample is transferred to the testing bin through a channel at the base of the vacuum container. The testing bin is equipped with three chambers, each pre-filled with Benedict's reagent for detecting carbohydrates. During testing, the reagent reacts with the soil sample, resulting in a color change that varies from blue to green, yellow, or brick red depending on the concentration of carbohydrates. The CAD model of the module attached to the rover chassis is seen in Figure 9 and The physical model of the module attached to the rover chassis is seen in Figure 10.

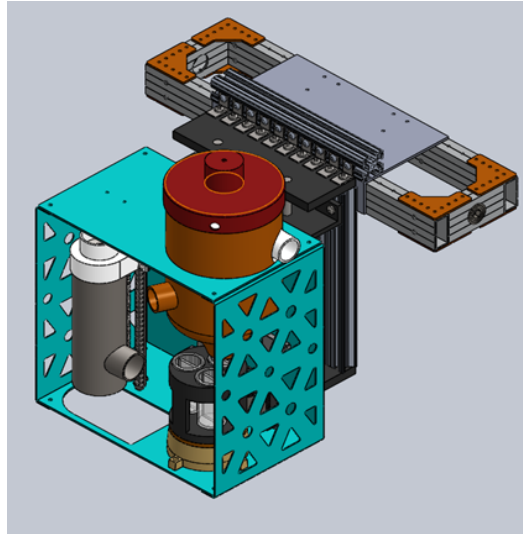

**Figure 9. Proof of life CAD model**

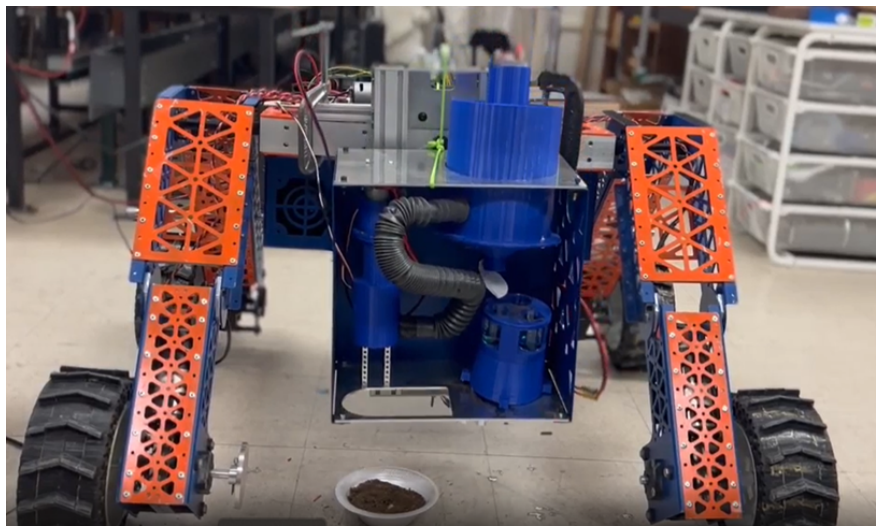

**Figure 10. Proof of life module physical hardware on rover chassis**

# IV. Results and Discussions

## A. Biochemical Testing of Soil Sample

*1. Protein Testing*

Proteins, composed of amino acids linked by peptide bonds, are key organic compounds indicative of life [9]. The Biuret test was used to detect proteins in the soil. Biuret reagent, blue due to copper (II) ions, reacts with peptide bonds to produce a deep blue or purple color. Approximately 2-3 grams of soil were mixed with 7 mL of diluted Biuret reagent. After mixing, the solution was allowed to decant for 2-3 minutes to observe the color change distinctly. Positive results for protein presence were identified by a deep blue color, confirming the reaction between the reagent and peptide bonds. Out of 10 soil samples tested, only 2 gave positive results for proteins, indicating their presence in limited locations.

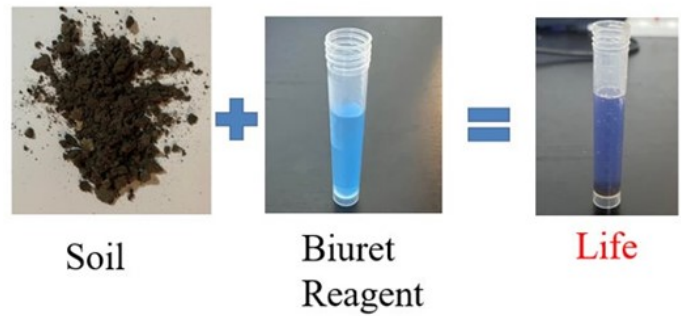

**Figure 11. Protein test**

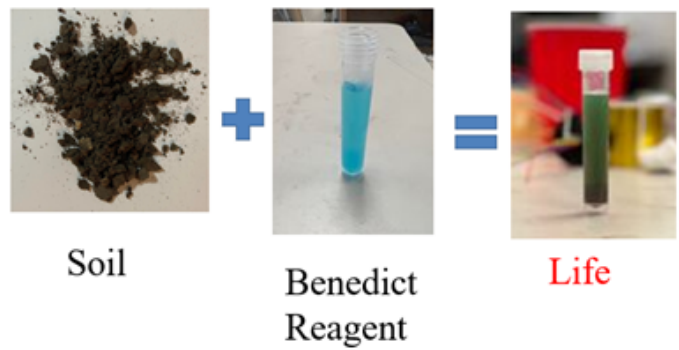

**Figure 12. Carbohydrate test**

*2. Carbohydrate Testing*

Carbohydrates, consisting of aldehyde (-CHO) and ketone groups, were tested using Benedict's reagent. The reagent, a mix of copper sulfate, sodium carbonate, and sodium citrate, reacts with reducing sugars in the soil to produce cuprous oxide, resulting in a color change. A 7 mL volume of diluted Benedict's reagent was combined with 2-3 grams of soil. The reaction produced color changes ranging from light green to brick red, depending on the concentration of carbohydrates. After allowing the solution to decant for 2-3 minutes, the color change became distinct and was captured by the camera. All 10 soil samples tested positive for carbohydrates, confirming their widespread presence.

## B. Post Processing

The camera mounted inside the module captured the color changes during the reactions, and the images were transmitted to a base station for further analysis. Carbohydrate tests were more reliable, as even trace amounts produced measurable reactions, whereas protein tests required a higher concentration of proteins to yield visible results. The higher probability of positive carbohydrate test results made it a more effective indicator for life detection. The system operated as designed, meeting all functional requirements and demonstrating its ability to detect life-related molecules. The findings confirmed the capability of the proof of life module to identify life-related organic compounds in collected soil samples.

## V. Conclusion and Future Work

The Mars Rover - Proof of Life module marks a significant advancement in autonomous life detection technologies for extraterrestrial exploration. By integrating mechanical excavation using different mechanisms in the module, onboard biochemical analysis, and autonomous control systems, the module presents a robust, self-contained approach to detect presence of life in planetary soil samples without the need for human intervention. The ability to excavate, collect, and test samples in situ represents a critical step toward enabling long-duration, self-reliant extraterrestrial missions. While the current system showcases prominent results, certain design limitations must be addressed in the future to enhance its scientific validity. The use of shared components such as the drill bit, vacuum system, and hose pipe, introduces the risk of cross-contamination during multi-site sampling. Future iterations should prioritize contamination control through modular components. Furthermore, the biochemical analysis is currently limited to protein and carbohydrate detection. Expanding the test suite to include a broader range of biosignature assays, supported by more sensitive and rapid detection methods, would significantly strengthen the system's ability to identify potential biological activity. Ultimately, this project serves as a proof of concept for autonomous life detection and lays the novel approach to detect the presence or absence of living components of life from soil samples for University Rover Challenge. With further refinement and innovation, such systems can become vital instruments aboard next-generation Mars rovers and deep space missions.